\documentclass[letterpaper,twocolumn,english,american,prl,showpacs]{revtex4}
\usepackage[T1]{fontenc}
\usepackage[latin1]{inputenc}
\usepackage{amsmath}
\usepackage{graphicx}
\usepackage{amssymb}

\makeatletter

\newcommand{\eps}{\varepsilon}
\newcommand{\mT}{-}

\newcommand{\mM}{-}
\newcommand{\pM}{+}
\newcommand{\PP}{P}

\usepackage{babel}

\begin{document}

\title{Designing the Dynamics of Spiking Neural Networks}

\author{Raoul-Martin Memmesheimer$^{1,2,3}$ and Marc Timme$^{1,2,4}$}

\affiliation{$^{1}$Max-Planck-Institut für Dynamik und Selbstorganisation (MPIDS),
37073 Göttingen, Germany,}

\affiliation{$^{2}$Bernstein Center for Computational Neuroscience (BCCN) Göttingen,}

\affiliation{$^{3}$Fakultät für Physik, Georg-August-Universität Göttingen,}

\affiliation{$^{4}$Center for Applied Mathematics, Cornell University, Ithaca,
NY 14853, USA}

\date{Tue Oct 24 09:46:03 CEST 2006
}

\begin{abstract}
Precise timing of spikes and temporal locking are key elements
of neural computation. Here we demonstrate how even strongly heterogeneous,
deterministic neural networks with delayed interactions and complex
topology can exhibit periodic patterns of spikes that are precisely
timed. We develop an analytical method to find the set of all networks
exhibiting a predefined pattern dynamics. Such patterns may be arbitrarily
long and of complicated temporal structure. We point out that the
same pattern can exist in very different networks and have different
stability properties. 
\end{abstract}


\maketitle
Repeated patterns of spikes with temporal precision in the millisecond
range have been experimentally observed in different neuronal systems
\cite{A93,S99,I04,GS05}. They correlate with internal and external
stimuli and are thus discussed to be essential for neural information
processing (see, e.g., \cite{A04}). Their dynamical origin, however,
is unknown. One possible explanation for their occurrence is the existence
of excitatorily coupled feed-forward structures, synfire chains \cite{A82,HHP95,DGA99},
which are embedded in a network of otherwise random connectivity and
receive a large number of random external inputs. Other studies point
out that spike patterns can originate as attractors of deterministic
recurrent networks if inhibitory interactions dominate \cite{J02,BR02}.
These studies could already treat networks of complicated connectivities and
successfully found one specific network solution for a given pattern.
Yet it is still unclear which set of networks have the potential to realize 
a given spiking dynamics. Moreover, the recent studies 
considered interactions without delays. Delays, however,
are known to be significant in biological neural systems \cite{EKKS91}
and to have a strong impact onto even the qualitative dynamics of
neural networks (cf.\ \cite{MS90,EPG95,TWG02a,RBH05,TWG02b,ZTGW04,DTDWG04}).

It is thus still an open question whether and how a deterministic network, despite simultaneously exhibiting delayed interactions and strong heterogeneities, can yet display precisely timed spiking dynamics. If so, what are 
the possible networks that generate a given dynamics?

In this Letter we study a class of spiking neural network models with
delayed interactions. We provide a solution to an 
inverse problem for networks of arbitrary connectivity: 
We present an exact analytical method to find
the set of all networks, by determining the coupling strengths, such
that they exhibit a given periodic spike pattern of arbitrary temporal
extent. 
The analysis shows that even arbitrarily large networks with
complicated connection topologies and strong heterogeneities can yet
display patterns of spikes that are timed precisely. The class of
networks realizing a simple periodic pattern, i.e.\ one in which
each neuron fires exactly once before the sequence repeats, is derived
and parameterized analytically. The network may have a mixture of
both excitatory and inhibitory couplings, with the stability of a
pattern depending on the particular coupling architecture.

\begin{figure}
\begin{center}\includegraphics[%
  width=88mm,
  keepaspectratio]{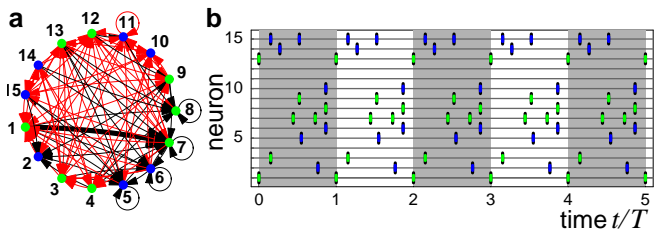}\end{center}
{~\vspace{-10mm}}
\caption{(color) Complicated spike pattern in a small network ($N=15$).
(a) Network of eight integrate-and-fire (green) and seven Mirollo-Strogatz
(blue) neurons with distributed thresholds $\Theta_{l}\in[0.5,2.0]$
and delays $\tau_{ml}\in[0.1,0.9]$. Each directed connection between
any two neurons is randomly chosen to be present with probability
$p=0.6$. Connections are either excitatory (black) or inhibitory
(red) (thicknesses proportional to coupling strengths). (b) The spiking
dynamics (green and blue bars according to neuron type) of the network
shown in (a) perfectly agrees with the predefined pattern (period
$T=1.3$) of precisely timed spikes (black bars underlying the colored
ones). The pattern includes several simultaneous spikes. Three neurons,
$l\in\{4,11,12\}$, are switched off (non-spiking). \label{cap:heterogeneity}}
\end{figure}

Consider a network of $N\in\mathbb{N}$ oscillatory neurons that interact by sending
and receiving spikes via directed delayed connections. One phase-like
variable $\phi_{l}(t)$ specifies the state of each neuron $l\in\{1,\ldots,N\}$
at time $t$. A strictly monotonic increasing rise function $U_{l}$
defines the membrane potential $U_{l}(\phi_{l})$ of the neuron, representing
its subthreshold dynamics \cite{MS90}. In the absence of interactions,
the phases increase uniformly obeying $d\phi_{l}/dt=1$. When $\phi_{l}$
reaches its threshold, $\phi_{l}(t^{-})=\Theta_{l}$, it is reset,
$\phi_{l}(t)=0$, and a spike is emitted. After a delay time $\tau_{ml}$
this spike signal reaches the post-synaptic neuron $m$, inducing
an instantaneous phase jump\begin{equation}
\phi_{m}\left(t+\tau_{ml}\right)=H_{\varepsilon_{ml}}^{(m)}\left(\phi_{m}\left(\left(t+\tau_{ml}\right)^{-}\right)\right),\label{eq:1}\end{equation}
 mediated by the transfer function $H_{\varepsilon}^{(m)}(\phi)=U_{m}^{-1}(U_{m}(\phi)+\varepsilon)$
that is strictly monotonic increasing both as a function of $\varepsilon$
and of $\phi$. Here, $\varepsilon_{ml}$ denotes the strength of
the coupling from neuron $l$ to $m$. Sending and receiving of spikes
are the only nonlinear events occurring in these systems. For simplicity
of presentation, we here focus on non-degenerate events: We consider
arbitrary periodic patterns in which (i) all spikes are sent at non-identical
times and (ii) received at non-identical times, and (iii) neurons
receiving a spike do not generate a new spike at the same time. We
focus on networks of identical neurons $U_{l}(\phi)\equiv U(\phi)$
with the same intrinsic inter-spike intervals fixed by $\Theta_{l}\equiv1$,
on identical delays $\tau_{ml}\equiv\tau$, and patterns
without silent  neurons (that do not spike within a pattern due to sufficiently strong inibitory input). Below, we will explain the underlying
ideas of how to find the set of all networks exhibiting a given pattern as an invariant solution for this class of systems. 
Nevertheless, based on the analysis presented
here, the developed method can be further extended \cite{MT06} to
cover also different types of neurons, heterogeneously distributed
delays and thresholds, and complicated stored patterns that include
degenerate spikes, multiple firings of the same neuron and silent
neurons that never fire. Figure \ref{cap:heterogeneity} illustrates
such a general case.

What characterizes a periodic pattern of precisely timed spikes? Let
$t_{i}$, $i\in\mathbb{Z}$, be an ordered list of times at which
a neuron emits the $i\textrm{th}$ spike occurring in the network,
such that $t_{j}>t_{i}$ if $j>i$. Assume a periodic pattern consists
of $M$ spikes. Such a pattern is then characterized by its period
$T$, by the times $t_{i}\in[0,T)$ of spikes $i\in\{1,...,M\}$,
and by the indices $s_{i}\in\{1,\ldots,N\}$ identifying the neuron
that spikes at $t_{i}\,$. To exclude technicalities in the presentation, we
assume that for all pairs $t_i$ and $t_j$ of subsequent spike times of each
neuron $l$, it receives at least one  spike within the interval
$(t_i,t_j)\cap(t_i,t_i+\Theta_l)$. Periodicity entails $t_{i}+nT=t_{i+nM}$
and $s_{i}=s_{i+nM}$ for all $n\in\mathbb{Z}$. This imposes conditions
on the time evolution of the neurons' phases. Suppose a specific neuron
$l$ fires at $K(l)$ times $t_{i_{k}}\in[0,T)$, $k\in\{1,...,K(l)\}$
within the first period. For the non-degenerate patterns considered, this
implies 
\begin{equation}
\phi_{l}(t_{i_{k}}^{-})=1,\label{eq:F1}
\end{equation}
 whereas at any other time $t\in[0,T)$, $t\neq t_{i_{k}}$ for all
$k$, \begin{equation}
\phi_{l}(t^{-})<1,\label{eq:S1}\end{equation}
 to prevent untimely firing. The monotonicity of the transfer function
implies that the periodicity of the pattern is necessary and sufficient
\cite{MT06} for the periodicity of the phases, \begin{equation}
\phi_{l}(t)=\phi_{l}(t+nT),\label{eq:P1}\end{equation}
for all $n\in\mathbb{Z}$ and all $t\in[0,T)$. We therefore equivalently
consider $\phi_{l}(t)$ for $t\in[0,T)$ with periodic boundary conditions.
All times are measured modulo $T$ and spike time labels $i$ are
reduced to $\{1,...,M\}$ by subtracting a suitable integer multiple
of $M$. Let $\PP(i)\in\{1,...,M\}$ denote the spike arriving last
before the firing time $t_{i}$ such that $\PP(i)=\text{argmin}\{ t_{i}-\theta_{j}\,|\, j\in\{1,...,M\}\}$,
where $\theta_{j}=t_{j}+\tau$ is the arrival time of the spike labeled
$j$.
\begin{figure}[h]
\centering
\includegraphics[width=55mm]{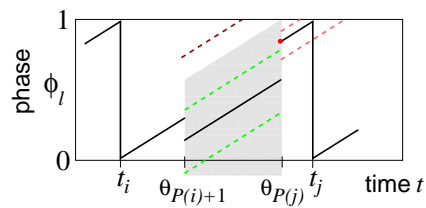}
{~\vspace{-2mm}}
\caption{(color) Restriction of a neuron's dynamics between its firing
events (\ref{eq:F2}). In this example, two spikes arrive between
the firing times $t_{i}$ and $t_{j}$ of neuron $l$. The solid line
indicates one possible time evolution of the phase $\phi_{l}(t)$.
Between the firing times, $\phi_{l}(t)$ may follow any path within
a possibly semi-infinite polygon (gray shaded; green dashed lines show other possible trajectories).
A too large phase at $\theta_{\PP(i)\pM1}$ contradicts (\ref{eq:F2})
and will lead to early firing (dark red dashed line). The phase at
$\theta_{\PP(j)}$ is fixed (red dot). Any other phase inconsistent
with the equality in (\ref{eq:F2}) would lead to a firing time earlier
or later than predefined (light red dashed lines). \label{cap:PatternN}}
\end{figure}

\begin{figure}
\begin{center}\includegraphics[%
  width=88mm,
  keepaspectratio]{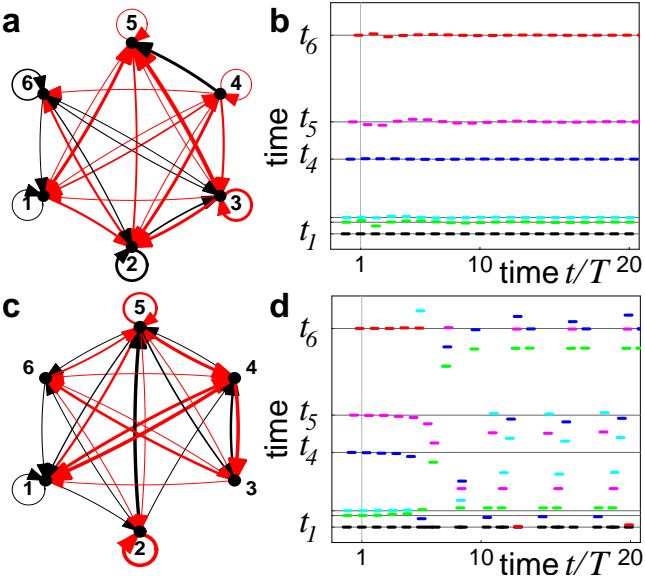}\end{center}

\caption{(color) Two different networks (a), (c) realize the same predefined
pattern ((b), (d) grey lines). A small random perturbation is applied
at the beginning of the second period. The network dynamics (spike
times relative to the spikes of neuron $l=1$, color coded for each
neuron), found by exact numerical integration
\cite{TWG03} shows that in network (a) the pattern is stable and thus regained
after a few periods (b); in network (c) it is unstable (d) and eventually
another pattern will be assumed. \label{cap:Single-periodic-patterns}}
\end{figure}

Moreover, let $\Delta_{j}=\theta_{j+1}-\theta_{j}$ be the time differences
between two successive arrivals. We can now rewrite Eqs.~(\ref{eq:F1})
and (\ref{eq:S1}) for neuron $l$ as a set of conditions on the phases
$\phi_{l}(\theta_{i})$ at each spike arrival time $\theta_{i}$,
in terms of the firing times $t_{i_{k}}$ of that neuron and spike
arrival times $\theta_{j}$,
\begin{align}
\phi_{l}(\theta_{\PP(i_{k})})= & 1-(t_{i_{k}}\mT\theta_{\PP(i_{k})}),\label{eq:F}\\
\phi_{l}(\theta_{j})< & 1-\Delta{}_{j},\label{eq:S}\end{align}
 where $k\in\{1,...,K(l)\}$ and $j\in\{1,...,M\}$, $j\neq\PP(i_{k})$
for all $k$. The coupling strengths $\eps_{ll'}$, $l,l'\in\{1,\ldots,N\}$
of a network realizing a given pattern are now restricted by a system
of $\sum_{l=1}^{N}K(l)=M$ nonlinear equations and $\sum_{l=1}^{N}(M-K(l))=(N-1)M$
inequalities originating from (\ref{eq:F}) and (\ref{eq:S}): After
a firing of neuron $l$ at time $t_{i}$ where its phase is zero,
conditions (\ref{eq:F}) and (\ref{eq:S}) impose restrictions at
each spike arrival time while the time evolution proceeds towards
the subsequent firing time $t_{j}$ of neuron $l$, as illustrated
in Fig.~\ref{cap:PatternN}. As a result, we have\begin{widetext}\begin{equation}
\begin{array}{rl}
H_{\eps_{ls_{\PP(i)\pM1}}}(\theta_{\PP(i)\pM1}-t_{i}) & <1-\Delta{}_{\PP(i)\pM1}\,,\\
H_{\eps_{ls_{\PP(i)\pM2}}}(H_{\eps_{ls_{\PP(i)\pM1}}}(\theta_{\PP(i)\pM1}-t_{i})+\Delta{}_{\PP(i)\pM1}) & <1-\Delta{}_{\PP(i)\pM2}\,,\\
 & \,\,\vdots\\
H_{\eps_{ls_{\PP(j)}}}(...H_{\eps_{ls_{\PP(i)\pM2}}}(H_{\eps_{ls_{\PP(i)\pM1}}}(\theta_{\PP(i)\pM1}-t_{i})+\Delta{}_{\PP(i)\pM1})\ldots+\Delta{}_{\PP(j)\mM1}) & =1-(t_{j}-\theta_{\PP(j)}).\end{array}\label{eq:F2}\end{equation}

\end{widetext} A particular solution \cite{MM03} to the system (\ref{eq:F2}), provides the coupling
strengths $\eps_{ll'}$, $l'\in\{1,\ldots,N\}$, of incoming connections
to neuron $l$. Solutions to systems analogous to (\ref{eq:F2}) for
all neurons $l$ define the coupling architecture of the entire network.
Often (\ref{eq:F2}) is an under-determined system such that many
solutions exist, implying that many different networks realize the
same predefined pattern; cf. Fig.\ \ref{cap:Single-periodic-patterns}.

We can then require additional properties from the network. For instance,
a connection from a neuron $l$ to $m$ can be absent (requiring the
coupling strength $\eps_{ml}=0$), taken to be inhibitory ($\eps_{ml}<0$),
excitatory ($\eps_{ml}>0$) or to lie within an interval. In particular,
we can specify inhibitory and excitatory subpopulations. In certain
cases, such as for networks of leaky integrate and fire neurons \cite{L07},
$U(\phi)=U_{\gamma}(\phi)=[1-\exp(-\gamma\phi)]/[1-\exp(-\gamma)]$,
$\gamma>0$, or Mirollo-Strogatz neurons \cite{MS90}, $U(\phi)=U_{b}(\phi)=b^{-1}\ln(1+(\exp(b)-1)\phi)$,
$b>0$, a solution of (\ref{eq:F2}) can be found in a simple way,
because the system is then reducible to be linear in the couplings
or polynomial in its exponentials.

Networks realizing a given pattern do not always exist. This can already
be observed from a simple example: Consider a pattern with no spike arrival
between two spikes sent by the same neuron. Due to its free evolution between
the spiking times, their time difference must equal the free period; hence a
predefined pattern with different inter-spike interval is not realizable by
any network.

For a simple periodic pattern, the system (\ref{eq:F2}) is guaranteed to have
a solution, as long as basic requirements (e.g., the delays being smaller than
the neurons' intrinsic inter-spike-intervals) are obeyed. Without loss of
generality the neuron firing at time $t_{l}$ is labeled $l$, i.e.\ $s_{l}=l$
for $l\in\{1,...,M\equiv N\}$. An analytic parameterization of all networks
realizing such a pattern is then given by \begin{align} \eps_{l\PP(l)\pM1}= &
H_{\phi_{l}(\theta_{\PP(l)\pM1})}^{-1}(\theta_{\PP(l)\pM1}\mT t_{l}),\nonumber
\\ \eps_{l\PP(l)\pM k}= & H_{\phi_{l}(\theta_{\PP(l)\pM
k})}^{-1}(\phi_{l}(\theta_{\PP(l)\pM k\mM1})+\Delta_{\PP(l)\pM
k\mM1}),\nonumber \\ \eps_{l\PP(l)}= &
H_{1-(t_{l}\mT\theta_{\PP(l)})}^{-1}(\phi_{l}(\theta_{\PP(l)\mM1})+\Delta_{\PP(l)\mM1}),\label{eq:epsi}\end{align}
in terms of the neurons' phases $\phi_{l}(\theta_{i})$, $l,i\in\{1,...,N\}$ at
the spike arrival times. Here $k\in\{2,...,N-1\}$ and $H_{\psi}^{-1}(\phi)$ is
the inverse of $H_{\eps}(\phi)$ with respect to $\eps$. The phases
$\phi_{l}(\theta_{i})$ are subject to the restrictions (\ref{eq:S}). This
parameterization shows that an $N(N-1)$-dimensional submanifold of networks
realizing the pattern exists in $\eps_{ij}$-space. Additional features entail
additional conditions on the phases at the spike arrival times: For instance,
exclusion of self-interaction is guaranteed by the conditions
$\phi_{l}(\theta_{l})=\tau$ if there is no spike-arrival in
$(t_{l},\theta_{l})$, and
$\phi_{l}(\theta_{l})-\phi_{l}(\theta_{l\mM1})=\Delta_{l\mM1}$ otherwise,
reducing the dimension of the submanifold of possible networks by $N$.
Moreover, requiring the couplings to be purely inhibitory leads to the
accessibility conditions
\begin{align}
\phi_{l}(\theta_{\PP(l)\pM1})\leq &
\theta_{\PP(l)\pM1}\mT t_{l},\label{eq:purelyinhibitory1}\\
\phi_{l}(\theta_{j\pM1})-\phi_{l}(\theta_{j})\leq &
\Delta_{j},\label{eq:purelyinhibitory2}
\end{align} 
where $j\neq\PP(l)$. We can therefore successively choose
$\phi_{l}(\theta_{\PP(l)\pM m})$, $m\in\{1,...,N-1\}$, starting with $m=1$.
Inequalities\ (\ref{eq:purelyinhibitory1}) and (\ref{eq:purelyinhibitory2})
hold with reversed relations for purely excitatory coupling. Purely inhibitory
realizations exist if a pattern has period $T>1$ (larger than the neuron's
intrinsic inter-spike interval); otherwise $\phi_{l}(t_{l}^{-})=1$ is not
accessible from $\phi_{l}(t_{l})=0$. Similarly, purely excitatory
realizations exist if a pattern has period $T<1$.

Is a pattern emerging in a heterogeneous network stable or unstable? We
numerically investigated patterns in a variety of networks and found that in
general the stability properties of a pattern depend on the details of the
network it is realized in, see Fig.\ \ref{cap:Single-periodic-patterns} for an
illustration. Depending on the network architecture, the same pattern can be
exponentially stable or unstable, or exhibit oscillatory stable or unstable
dynamics. For any specific pattern in any specific network, the linear
stability properties can also be determined analytically, similar to the exact
perturbation analyses for much simpler dynamics in more homogeneous networks
\cite{TWG02a,TWG02b}. More generally, in every network of neurons with
congenerically curved rise functions and with purely inhibitory (or purely
excitatory) coupling, a nonlinear stability analysis shows that the possible
non-degenerate patterns are either \emph{all} stable or \emph{all} unstable.
For instance, in purely inhibitory networks of neurons with rise functions of
negative curvature, such as integrate-and-fire neurons, every periodic
non-degenerate spike pattern, no matter how complicated, is stable.

In summary, we presented a method to find the set of all networks realizing a
predefined periodic pattern of spikes, and for imposing additional
constraints, for instance specifying absent connections and choosing
inhibitory or excitatory subpopulations. A predefined simple periodic pattern
is particularly interesting because a network realizing it is guaranteed to
exist; here we parameterized analytically all such networks.

In general, these results demonstrate that precise, reproducible dynamics
arises even in high-dimensional heterogeneous complex systems where it
might be unexpected. The design method of solving an inverse problem presented
here, capable of finding all networks that exhibit a predefined dynamics,
might thus be of interest in the theory of coupled oscillators and complex
networks and contributes a novel perspective to theoretical neuroscience. The
method can be extended to include heterogeneities in all parameters and
non-degenerate patterns \cite{MT06}, hidden neurons outside a core network, as
well as non-periodic patterns, by dropping the periodicity constraint
(\ref{eq:P1}).

In particular, our results shed a new light on how patterns of precisely timed
spikes may emerge in deterministic neural network dynamical systems, even for
biologically realistic architecture. For instance these networks may
simultaneously exhibit strongly heterogeneous parameters, complicated
topology, and substantial and distributed delays. However, future work still
needs to fully answer how experimentally observed synchronization \cite{S99}
as well as recurrent patterns of spikes \cite{A93,I04,GS05} really arise.
Important topics of theoretical research include, for instance, (i) the exact
dynamics of networks of excitable neurons that are either excited by recurrent
network inputs or by external stimuli \cite{HM01}, (ii) spike patterns that
are not periodic but separated by intervals of irregular activity \cite{MPF05}, as well as
(iii) cortical songs which are sequences of patterns occurring repeatedly in
the same order but varying in the timing between patterns \cite{I04,GS05}.
Studies in these directions could further clarify mechanisms used for
information processing in such networks. In particular, it will be exciting to
see whether biological neural networks rely on stochastic features of topology
and input \cite{DGA99} or the exact wiring diagram and precise single neuron
dynamics play a significant role in creating a temporal code.

We thank S.\ Jahnke, F.\ Wolf and T.\ Geisel for useful discussions
and the BMBF Germany for partial support under grant number 01GQ0430.

\bibliographystyle{apsrev}

\end{document}